\documentclass[proof]{WileyASNA-v1}
    
\articletype{Article Type}%
    
\newcommand{\rin}{r_{\mathrm{in}}}
\newcommand{\risco}{r_{\mathrm{isco}}}

\begin{document}
    
\title{Spectra of Puffy Accretion Discs: the \texttt{kynbb} Fit}
    
\author[1,2]{Debora Lan\v{c}ov\'{a}*}
\author[3,4]{Anastasiya Yilmaz}
\author[5]{Maciek Wielgus}
\author[3]{Michal Dov\v{c}iak}
\author[6,7]{Odele Straub}
\author[1]{Gabriel Török}
    
\authormark{ \textsc{Lan\v{c}ov\'{a} et al}}
    
\address[1]{\orgdiv{Institute of Physics}, \orgname{Silesian University in Opava}, \orgaddress{\country{Czech Republic}}}
\address[2]{\orgdiv{N. Copernicus Astronomical Centre} \orgname{Polish Academy of Sciences}, \orgaddress{\state{Warsaw}, \country{Poland}}}
\address[3]{\orgdiv{Astronomical Institute}, \orgname{Czech Academy of Sciences}, \orgaddress{\state{Prague}, \country{Czech Republic}}}
\address[4]{\orgdiv{Astronomical Institute}, \orgname{Charles University}, \orgaddress{\state{Prague}, \country{Czech Republic}}}
\address[5]{\orgdiv{Max-Planck-Institut f\"ur Radioastronomie}, \orgaddress{\state{Bonn}, \country{Germany}}}
\address[6]{\orgdiv{ORIGINS Excellence Cluster}, \orgaddress{\state{Garching}, \country{Germany}}}
\address[7]{\orgdiv{Max-Planck-Institut f\"ur extraterrestrische Physik},  \state{Garching}, \country{Germany}}
    
\presentaddress{}
    
\corres{*Debora Lan\v{c}ov\'{a}. \email{debora.lancova@physics.slu.cz}}

\abstract{Puffy disc is a~numerical model, expected to capture the properties of the accretion flow in X-ray black hole binaries in the luminous, mildly sub-Eddington state. We fit the \texttt{kerrbb} and \texttt{kynbb} spectral models in \textsc{xspec} to synthetic spectra of puffy accretion discs, obtained in general relativistic radiative magnetohydrodynamic simulations, to see if they correctly recover the black hole spin and mass accretion rate assumed in the numerical simulation. We conclude that neither of the two models is capable of correctly recovering the puffy disc parameters, which highlights a~necessity to develop new, more accurate spectral models for the luminous regime of accretion in X-ray black hole binaries. We propose that such spectral models should be based on the results of numerical simulations of accretion.}
     
    
\jnlcitation{\cname{
\author{Lan\v{c}ov\'{a} D.}, 
\author{Yilmaz A.}, 
\author{Wielgus M.}, 
\author{Straub O.}, and 
\author{Török G.}} (\cyear{2022}), 
\ctitle{Spectra of puffy accretion discs: the \texttt{kynbb} fit}, %\cjournal{Q.J.R. Meteorol. Soc.}, \cvol{2017;00:1--6}.}
\cjournal{AN}, \cvol{2022}.}
    
\maketitle
    
\section{Introduction}\label{sec1}
    
Black hole (BH) accretion is commonly described using a~geometrically thin and optically thick disc model \citep{Shakura1973, Novikow1973}. This simple analytic framework proved to be extremely useful and capable of reproducing observed thermal spectra of X-ray black hole binary (XRB) systems with luminosities $L \sim 0.1\, L_{\rm Edd} < 0.3\, L_{\rm Edd}$, where $L_{\rm Edd}$ is the Eddington luminosity \citep{McClintock2014}. Nevertheless, the basic analytic models predict violent instabilities in the radiation pressure-dominated regime \citep{thermal76}, which is inconsistent with observations. These issues are addressed in the numerical magnetohydrodynamic models, where thermal and viscous stability appear to be a~consequence of the pressure support from the magnetic field \citep{Sadowski2016}. The corresponding numerical models, referred to as puffy discs \citep{Lancova2019}, are computationally very expensive, precluding detailed mapping of spectral properties across the model parameter space. Hence, it is interesting to quantify errors and biases of existing spectral fitting tools, when interpreting synthetic spectra of puffy discs as a~proxy of luminous XRB system with $L \gtrapprox 0.3\, L_{\rm Edd}$ for which geometrically thin disc approximation fails. In \citet{Wielgus2022} we have demonstrated biases of the {\tt kerrbb} (\citealt{Li05}, incorporating a~thin disc of \citealt{Novikow1973}) and {\tt slimbh} (\citealt{Straub2011}, incorporating a~slim disc of \citealt{Abramowicz1988}) spectral models. In this work, we test the {\tt kynbb} model \citep{Dovciak04} and compare its performance with that of {\tt kerrbb}.
    
\section{The puffy disc model}\label{sec2}
    
The puffy disc is a~model of accretion based on numerical General Relativistic Radiative Magnetohydrodynamic (GRRMHD) simulations of plasma infalling onto a~stellar mass BH. The simulations were performed using the {\texttt{KORAL}} code \citep{Sadowski2013, Sadowski2014, Sadowski2017}, which evolves the general relativistic equations for conservation of mass, momentum, and energy on a~logarithmic 3D spherical grid, under ideal MHD approximation. The radiation field is evolved using the M1 closure scheme \citep{Levermore1984,Sadowski2013}, an approximation  assuming existence of a~frame of reference in which radiation is isotropic. The numerical GRRMHD framework allows for a~physically self-consistent treatment of accretion with the presence of a~magnetic field, dynamically important radiation, and effects such as turbulent transport of angular momentum, advection, and outflows. Hence, GRRMHD simulations may provide a~reliable model of accretion disc spectra even in the luminous regime, in which analytic models fail to reproduce observations.
    
The puffy disc is a result of the simulation of accretion flow in the innermost region, where radiation pressure strongly dominates over gas pressure, in a~mildly sub-Eddington regime, in which thermal stability is supported by the magnetic pressure \citep{Sadowski2016}. Our GRRMHD simulations were performed for a~non-spinning (spin $a_*=0$) stellar mass BH of $M = 10\,M_{\odot}$ (where $M_{\odot}$ is the solar mass). The particular simulation discussed in this work has a~mass-accretion rate $\dot{M} \approx 0.6 \,\dot{M}_{\rm Edd}$\footnote{Here $\dot{M}_{\rm Edd} = L_{\rm Edd}/(\eta c^2)$ is defined using the thin disc efficiency $\eta = 0.057$ and the Eddington luminosity $L_{\rm Edd} = 1.25\cdot10^{38}M/M_\odot\,\mathrm{erg\cdot s^{-1}}$. Throughout this work, we are using the gravitational radius $r_g=GM/c^2$ as a~unit of length.}. 
    
\begin{figure}
    \centering
    \includegraphics[width=\linewidth]{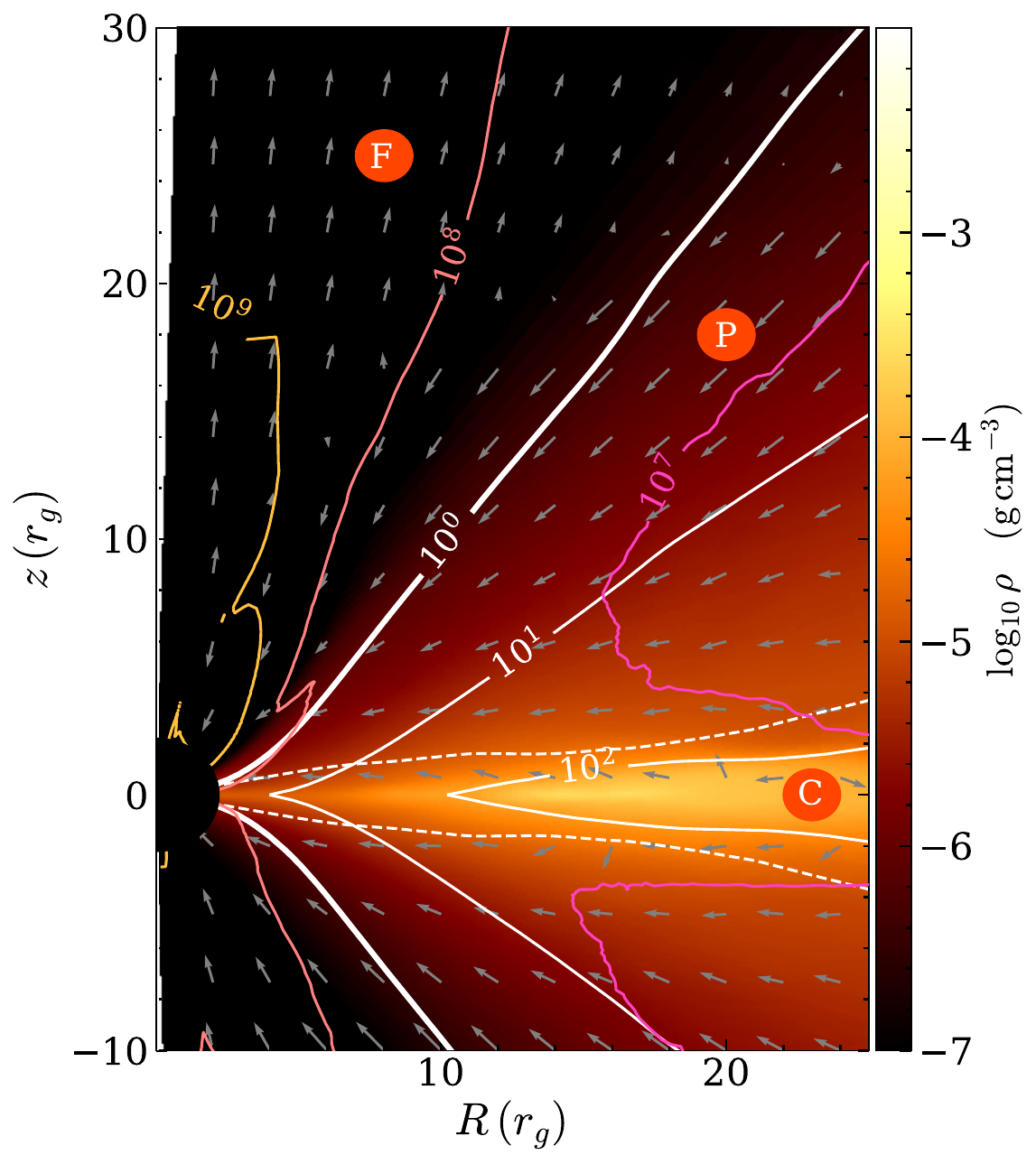}
    \caption{The cross-section of a~puffy disc averaged over time and azimuthal angle, showing the color map of density in the background, clearly separating the high-density core (C), the puffy region (P) and the funnel (F). Grey arrows indicate the gas momentum flow, the white continuous lines are the isolines of optical depth $\tau$ (the photosphere is located at $\tau=1$) and the white dashed lines near the equatorial plane show the density scale-height of the disc. Pink, orange and yellow contours are the gas temperature lines of $10^7,\,10^8$, and $10^9$ K ($0.9,\,8.6$, and $86.7$ keV) respectively.}
    \label{fig:puf} 
\end{figure}

The puffy disc model is specific for its vertical structure separating a~geometrically thin dense core, which resembles a~standard thin disc, and a~geometrically thick puffy region sandwiching the core, see FIGURE \ref{fig:puf}. The puffy region, distinguishing this model from the analytic solutions, is mildly optically thick, strongly magnetized, and extends up to the height of the elevated photosphere. The puffy region contributes significantly to the total mass accretion rate in the system. The funnel region near the axis is optically thin, and filled with hot and rarefied out-flowing plasma.
    
\subsection{Synthetic Observations of the Puffy Disc}
    
The elevated optically thick photosphere and magnetized puffy region roughly resemble a~warm slab corona \citep{Zhang2000,Gronki2020}, distinguish puffy discs from analytic models in the sub-Eddington regime and influence the observable features of the system by scattering the photons emitted in the core region. The dominant effect on the apparent observed luminosity, absent in the standard thin disc models, is the obscuration of the innermost region by the geometrically thick puffy region at high inclinations. At low inclinations, on the other hand, the puffy region collimates the radiation, leading to super-Eddington apparent isotropic luminosities \citep{Wielgus2022}.
    
In order to obtain synthetic spectra we post-processed results of the GRRMHD simulations using the \texttt{HEROIC} code \citep{Zhu2015,Narayan2016}, which solves a~complete radiative transfer problem on a~time-averaged simulation output. We created \texttt{.fits} files with the synthetic puffy disc spectra, assuming a~distance of $10 \,\mathrm{kpc}$ and $10\,\%$ uncertainties in the measured energy distribution. In this work we used the synthetic data to simulate an observation and fit the data using the \texttt{fakeit} function in \textsc{xspec}. We then used the \texttt{kerrbb} and \texttt{kynbb} to fit the simulated observations and obtain the estimates of the system parameters. 
    
A detailed description of the puffy disc GRRMHD model and its properties can be found in \cite{Sadowski2016} and \cite{Lancova2019}. The discussion of the observational features of puffy discs and details of the procedures employed to interpret synthetic observations within the \textsc{xspec} framework are given in \cite{Wielgus2022}.
    
\section{Spectral Modeling and Analysis}\label{spectral modeling and analysis}

    \subsection{Spectral Modeling}\label{modeling}
    
The \texttt{kerrbb} spectral model \citep{Li05} describes the spectrum of a~geometrically thin and optically thick Keplerian disc around a~Kerr BH, employing curved spacetime ray-tracing and accounting for the relativistic effects (light bending, self-irradiation or returning radiation, gravitational redshift, frame dragging, and Doppler boost) on the observed spectrum. The total power is calculated as a~sum over black-body annuli with the radial temperature profile following \citet{Novikow1973}. The model assumes the inner edge of the accretion disc located at the innermost stable circular orbit $\risco$ ($\risco=6\,r_g$ for the Schwarzschild case, and decreases with the BH spin). 
    
Similar to \texttt{kerrbb}, \texttt{kynbb} also relies on the ray-tracing algorithms to compute the transfer functions for the tables used in the model for photon paths in Kerr space-time, describing a~\citet{Novikow1973} accretion disc. It is a~local model developed as an extension to the \texttt{KY}\footnote{\url{https://projects.asu.cas.cz/stronggravity/kyn/}} package by \citet{Dovciak04}. In contrast to \texttt{kerrbb}, \texttt{kynbb} offers an option to define the inner disc radius $\rin$ either as a~free parameter or fixed at a~certain value other than $\risco$, computing the spectra starting from either $\risco$ or a~preferred $\rin$ value. Hence, \texttt{kynbb} model can produce spectra of a~truncated disc. For this work, we assumed $\rin \ge \risco$. One can choose to calculate the spectrum only for a~range of disc radii. Additionally, one also has the freedom of including polarization calculations but it is not used in this work. In addition to the thermal disc component, for both \texttt{kerrbb} and \texttt{kynbb} fits we include the \texttt{nthcomp} thermal Comptonization model to account for the up-scattering of the accretion disc seed photons to higher energies \citep{Zdziarski1996,Zycki1999} in the puffy region, see FIGURE~\ref{fig:puf}.

\begin{figure}[h!]
    \includegraphics[width=\linewidth]{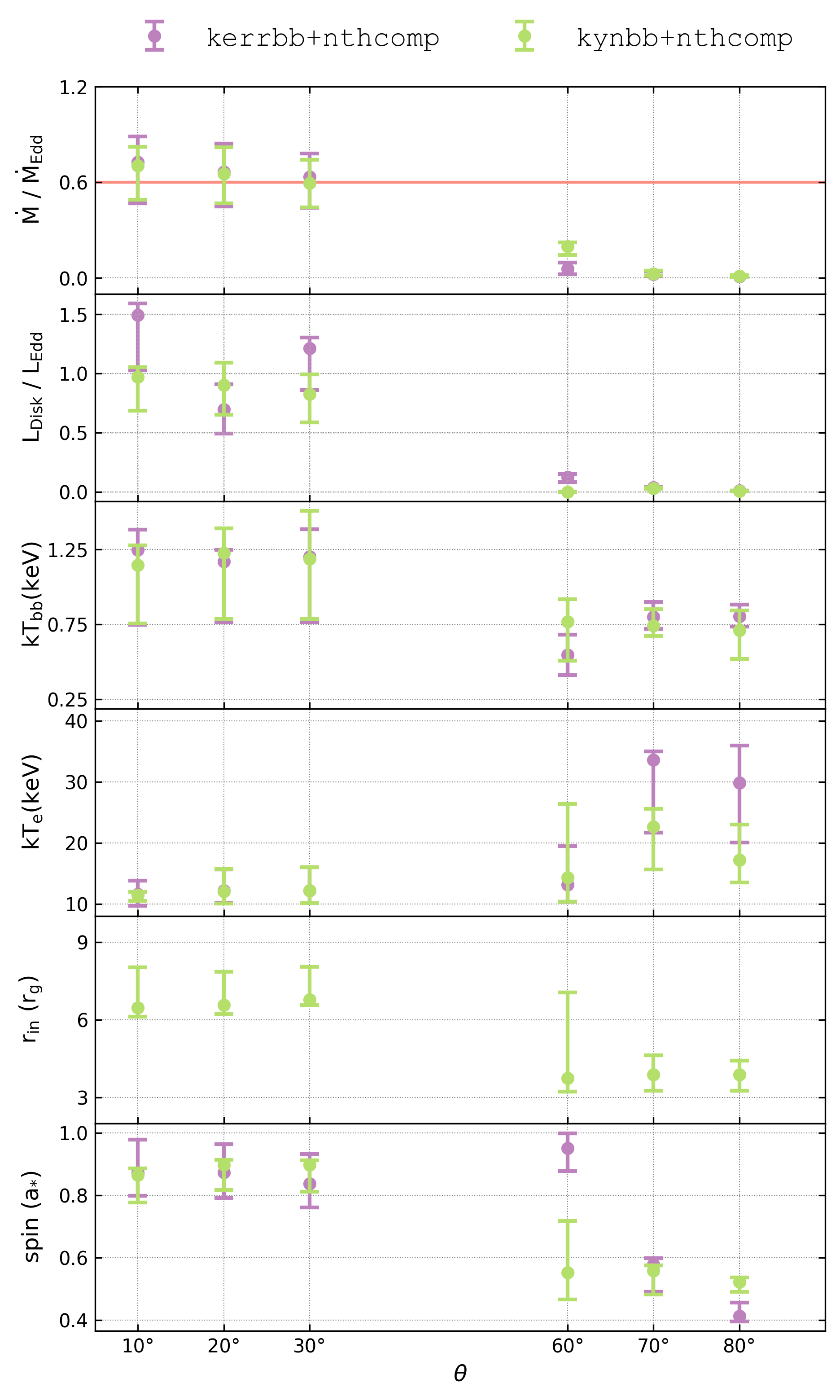}
    \caption{Parameter values obtained from spectral fitting with \texttt{kerrbb} (purple) with free ${a_{*}}$ and ${\dot{M}}$ and \texttt{kynbb} (green) with free ${\rin}$, ${a_{*}}$ and ${\dot{M}}$ as a~function of the observing inclination angle $\theta$. The pink solid line on the top panel highlights the assumed ${\dot{M}=0.6 \; \dot{M}_{\mathrm{Edd}}}$ in the GRRMHD simulations. The true assumed spin in the GRRMHD simulation is $a_* = 0$. }
    \label{fig:free_all}    
\end{figure}
    
\subsection{Spectral Analysis Results}\label{spectral_analysis}
    
For the analysis of the synthetic puffy disc spectra the \texttt{HEASOFT} (v.6.29), package \textsc{xspec} (v.12.12) \citep{Arnaud96} and \texttt{PyXspec}\footnote{\url{https://heasarc.gsfc.nasa.gov/xanadu/xspec/python/html/}}, a~Python interface to the \textsc{xspec} is used. Each spectra are generated for inclination angles of $\theta = 10^{\circ},\, 20^{\circ},\, 30^{\circ},\, 60^{\circ},\, 70^{\circ}$, and $80^{\circ}$ measured from the rotational axis.
    
Throughout the analysis, the BH mass, inclination angle, and distance (normalization parameter of \texttt{kynbb}) are fixed to their respective values used in the GRRMHD simulations while electron and seed photon temperatures ($\mathrm{T_{e}}$ and $\mathrm{T_{bb}}$, respectively) and the photon index $\mathrm{\Gamma}$ in \texttt{nthcomp} are set free to vary. For simplicity, we refer to each model (\texttt{kerrbb+nthcomp} and \texttt{kynbb+nthcomp}) by the specific model of the thermal component. With each model, the convolution model \texttt{cflux} in \textsc{xspec} is used to calculate $L_{\mathrm{Disc}}/L_{\mathrm{Edd}}$ for the energy range of 0.01-100.0 keV.
    
\begin{figure}[ht!]
    \includegraphics[width=\linewidth]{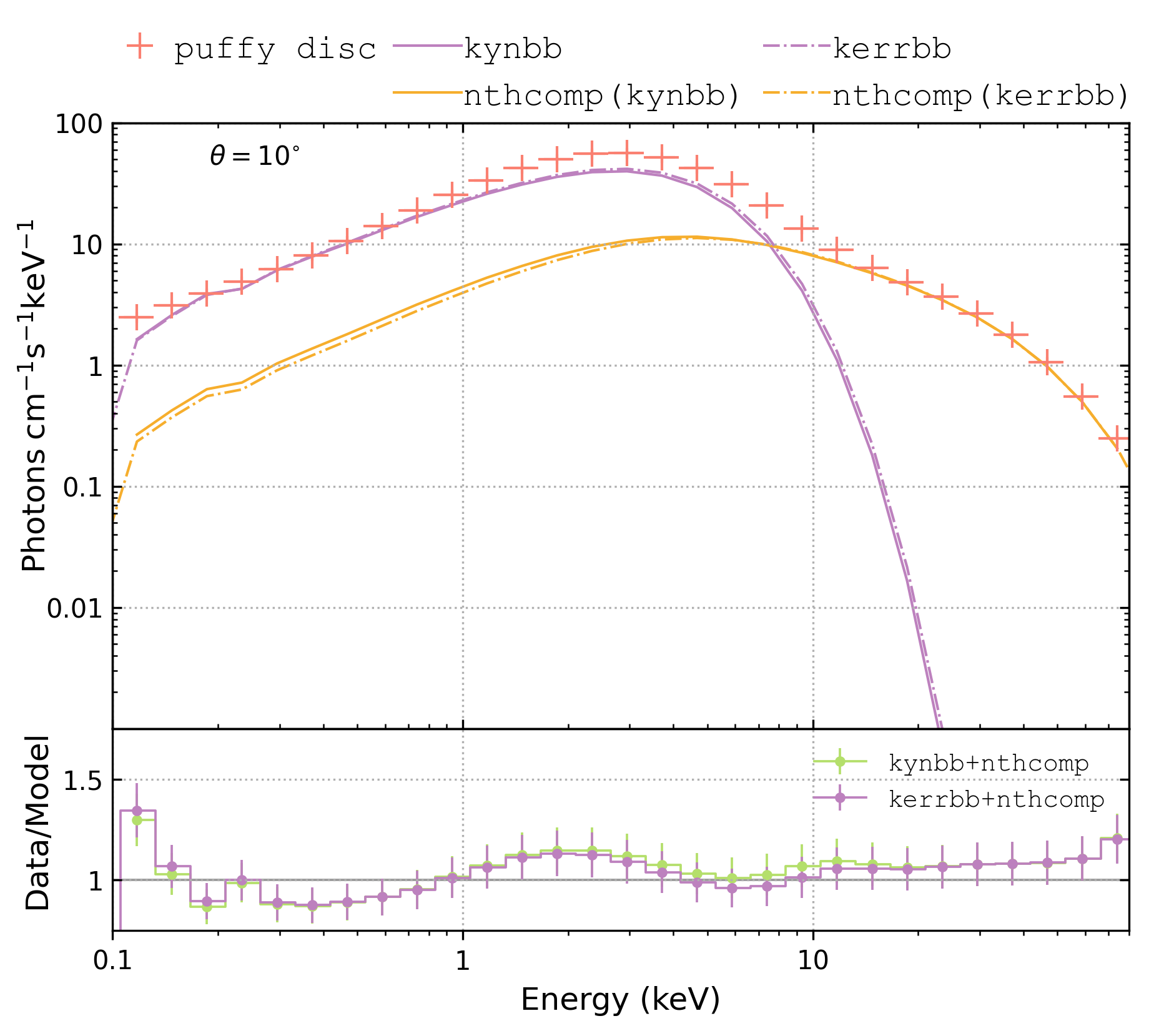}
    \includegraphics[width=\linewidth]{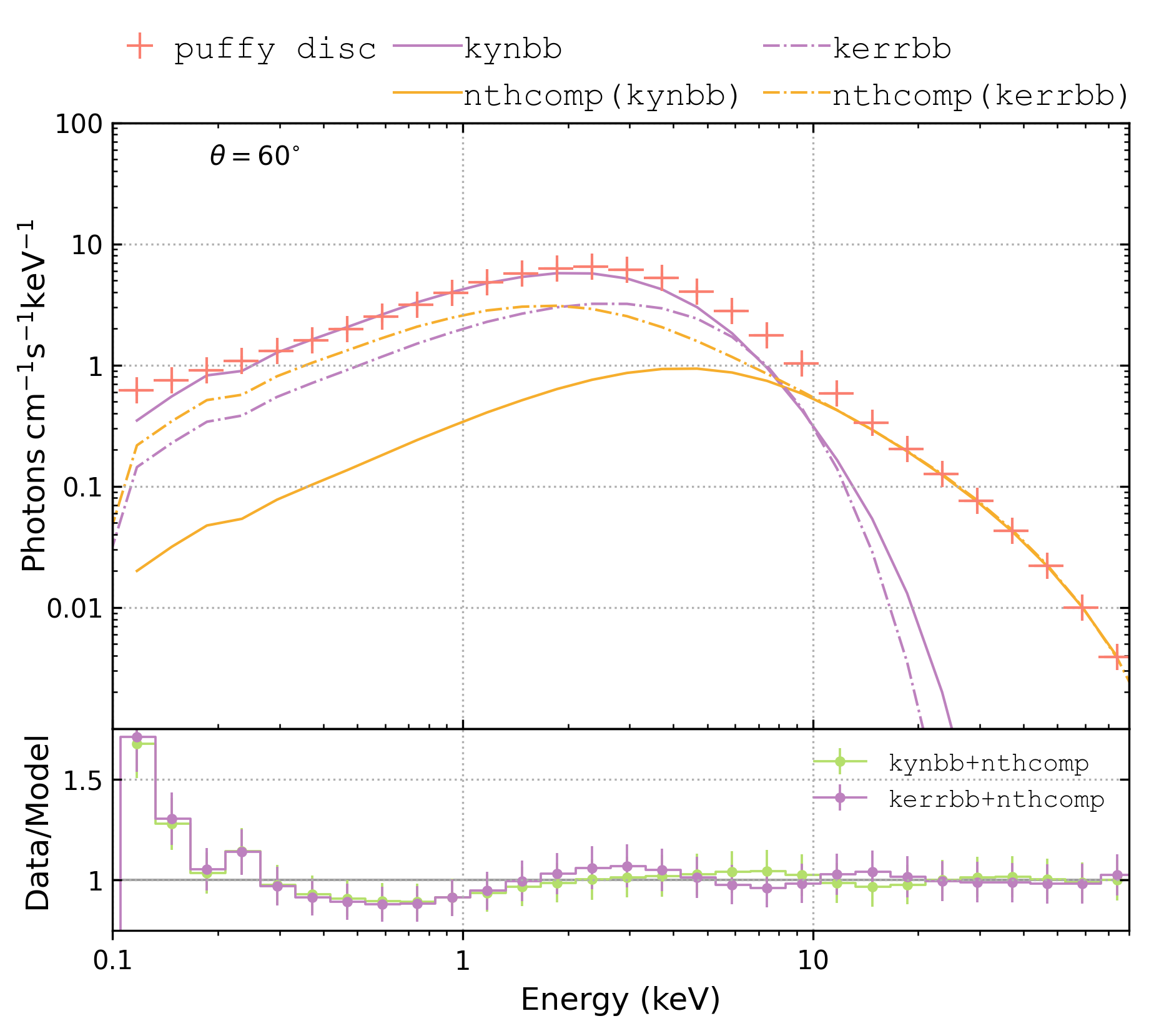}
    \caption{Synthetic puffy disc spectra (red crosses) calculated for the inclination angles of $\theta = 10^{\circ}$ and $\theta = 60^{\circ}$, fitted by the two spectral models with components indicated separately (\texttt{nthcomp}: orange, thermal disc component: purple). Dashed lines correspond to the \texttt{kerrbb} model setup while solid lines denote the \texttt{kynbb} model. Residuals between the fitted models and synthetic spectra of the puffy disc are also shown.
    \textit{Top}: Free ${a_{*} }$ and ${\dot{M}}$ in \texttt{kerrbb} and free ${\rin}$, ${a_{*}}$, and ${\dot{M}}$ in \texttt{kynbb} for inclination $\theta = 10^{\circ}$, \textit{Bottom}: Free ${a_{*} }$ and ${\dot{M}}$ in \texttt{kerrbb} and free ${\rin}$, ${a_{*}}$, and ${\dot{M}}$ in \texttt{kynbb} for $\theta = 60^{\circ}$.} 
    \label{fig:spec_10}    
\end{figure}

First, the spectra are analyzed using \texttt{kerrbb}, with BH spin $a_*$ and  $\dot{M}$ as free parameters additional to the \texttt{nthcomp} parameters. We then replaced \texttt{kerrbb} by \texttt{kynbb} with ${\rin}$ as an additional free parameter representing the inner edge of the disc. Low inclination angles produced reduced $\chi^{2}$ ($\chi^{2}/\rm{d.o.f.}$) in the range ${0.55 < \chi^{2}/\rm{d.o.f.}< 0.92}$ with \texttt{kerrbb} and ${1.28 < \chi^{2}/\rm{d.o.f.}< 1.37}$ with \texttt{kynbb} while the obtained values exceeded 2.0 for higher inclination angles, with larger residuals observed in both models. We present the fitted values of parameters obtained from \texttt{kerrbb} with free $a_*$ and $\dot{M}$ in comparison with additional free ${\rin}$ in FIGURE \ref{fig:free_all} and the synthetic spectra and residuals in ratios obtained from the fitting procedure with both models in the top panel of FIGURE~\ref{fig:spec_10} for $\theta=10^{\circ}$ and bottom panel for $\theta=60^{\circ}$.

\subsection{Effect of the Disc Inner Edge Position}
    
The second panel from the bottom in FIGURE~\ref{fig:free_all} shows the obtained value of ${\rin}$ for varying inclination angle $\theta$ assumed for the calculations. At low inclination the \texttt{kynbb} model indicates a~preference for a~truncated disc, with ${\rin} > 6 r_{\rm g}$, while the fitted spin values are large and positive. 
   On the other hand, for higher inclinations, ${\rin}$ moves closer to the ${\risco}$ corresponding to the obtained $a_*$, resulting in higher energy of the thermal component. With higher inclination, the \texttt{kerrbb} thermal component  has to be compensated by stronger \texttt{nthcompt} component even in the low energy bands, however, in the \texttt{kynbb} case, the thermal component is reproducing the thermal part of the puffy disc spectra well and Comptonization is prominent only in the higher energy bands. This provides more consistent (yet still inaccurate) spin outputs than the \texttt{kerrbb} case, see the bottom panel of FIGURE \ref{fig:free_all}.    
    
\begin{figure}
    \centering
    \includegraphics[width=\linewidth]{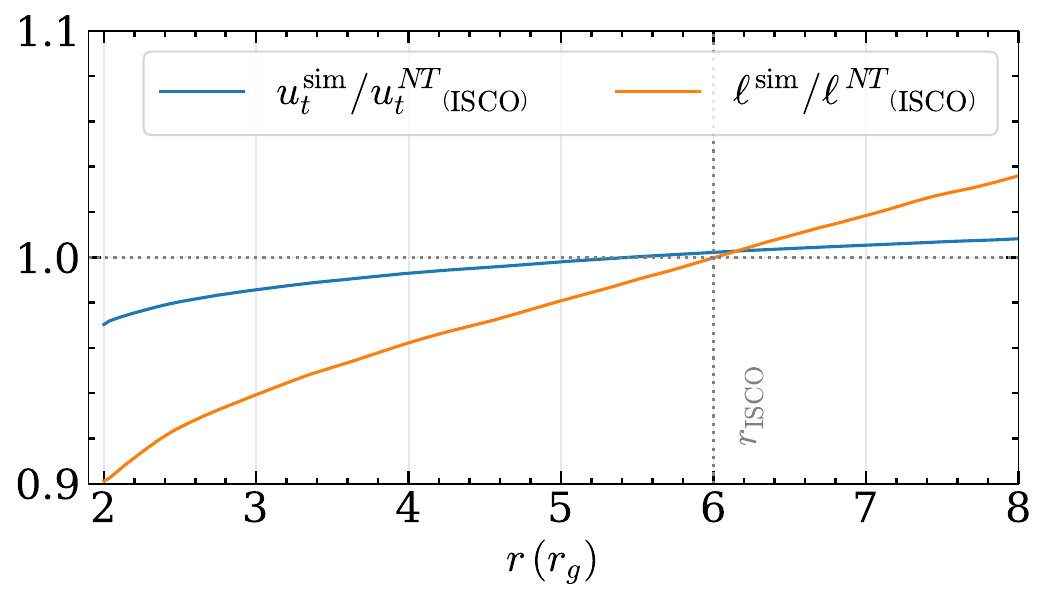}
    \caption{Values of the time component of gas four-velocity ($u_t^{\mathrm{sim}}$) and angular momentum ($\ell^{\mathrm{sim}}$) on the equatorial plane from the simulation compared to the Novikov-Thorne (NT) disc values at $\risco$.}
    \label{fig:toisco}
\end{figure}
    
When tested with spectra obtained from RXTE (Rossi X-Ray Timing Explorer) observations of GRO J1655-40 (Yilmaz et al., in preparation), both \texttt{kerrbb} and \texttt{kynbb} performed almost identically in measuring the spin and disc temperature when ${\rin}={\risco}$ is assumed in both models. There is a~significant mismatch between the results when the ${\rin}$ is set as a~free parameter in \texttt{kynbb} for the same mass accretion rates, due to the non-zero torque assumption at ${\rin} \neq {\risco}$. This assumption becomes much more important as one approaches the ISCO and eventually reaches the horizon. While this does not produce a~significant difference in model parameters in FIGURE~\ref{fig:free_all} and model components in FIGURE~\ref{fig:spec_10} for lower inclination angles, there is a~slight shift in the thermal component of the continuum to higher energies for higher inclination angles. This can be explained with parameter degeneracy between ${\rin}$ and $a_*$. For larger inclinations, it becomes more difficult to successfully model an obscured thermal component using standard relativistic thin accretion disc models. While the obtained spin values were significantly lower for higher inclination angles, ${\rin \; \sim 3 \; r_{g}}$ (FIGURE~\ref{fig:free_all}) corresponds to a~thermal component tail in \texttt{kynbb} extending to higher energies compared to \texttt{kerrbb} due to increased radiative efficiency of the accretion disc. 
  
\section{Conclusions}\label{sec5}   
    
Spin measurements via the X-ray spectral continuum fitting rely on accurately determining the position of the innermost edge of the accretion disc, assuming it terminates at the ISCO \citep{McClintock2014}. How accurately this can be achieved has an impact on the radiative efficiency of the disc in converting the accreted mass into outgoing radiation \citep{Li05}. While this radius assumption is at the core of the spectral continuum fitting method for spin measurements, it is found to cause difficulties in fitting procedures and discrepancies in model parameters when that assumption might not hold for all XRBs (Yilmaz et al., in preparation). \texttt{kynbb} with a~varying $\rin$ implies a~non-zero torque at the inner edge of the accretion disc, a~more realistic picture where magnetic fields that are strong enough to exert such torques might exist \citep{Gammie99, Krolik99}. This, in fact, results in a~significantly increased efficiency. While setting ${\rin}$ as a~free parameter in \texttt{kynbb} and computing the spectra from that certain value allows one to compare the effect of ${\rin}={\risco}$ assumption on the measured spin, these two parameters are degenerate when both are set free, resulting in consistent spin values when compared to \texttt{kerrbb} (see the bottom panel of FIGURE \ref{fig:free_all}). 
    
The puffy disc {numerical GRRMHD model does not terminate at the ISCO, and it does not have a~well-defined inner edge.
Below the ISCO matter is almost freely falling onto the BH, filling up the plunging region with hot, but still optically thick material, which contributes to the observable signal. This is in accordance with the technique of how the plunging region is modeled in \texttt{kynbb} - the fluid below ISCO is assumed to be free-falling and has the same energy and angular momentum as the matter which is orbiting at the ISCO. In the puffy disc simulation, the fluid behaves similarly -- the energy and angular momentum under the ISCO is almost the same as on ISCO, where the values coincide with the analytical values on the inner edge of the Novikov-Thorne disc, see FIGURE \ref{fig:toisco}. For both \texttt{kerrbb} and \texttt{kynbb} (and similarly also for \texttt{slimbh}), the plunging region does not correspond to the accretion disc structure with physically reliable parameter values. Successful modeling of this region requires a~more careful treatment of the accretion disc structure, which cannot be achieved with the traditional relativistic thin accretion disc models such as \texttt{kerrbb} and \texttt{kynbb}. In this paper we investigated whether there exists an \textit{effective} $r_{\rm in}$ that would allow to interpret puffy disc spectra within the \texttt{kynbb} model framework, correctly estimating the BH spin parameter. Based on our findings the answer to this question appears to be negative.

Our conclusions are consistent with those recently found by \citet{Wielgus2022} -- the available spectral fitting models fail to correctly interpret synthetic spectra of puffy discs and to recover the assumed value of the BH spin. In order to address these issues a~new spectral model, using results obtained from GRRMHD simulations, and interpolating on a~grid of numerical models, should be developed. A complete analysis should necessarily involve simulation of puffy discs around BHs with non-zero spin. However, the progress is limited by the immense computational costs of the GRRMHD simulations ($\sim10^7$ CPU hours). Hence, we have focused so far on the stability considerations \citep{Sadowski2016} and the comparison of different mass accretion rates \citep{Lancova2019}, leaving investigations of the spin impact for the future work. A~spectral model mapped on GRRMHD simulations across varying system parameters, including spin, could enable us to extend the continuum fitting method to more luminous sources and thus provide new valuable insight into studies of luminous XRBs.

\section*{Acknowledgments}
    
We thank M. Abramowicz and W. Kluzniak for fruitful discussions. G.T. and M. D. acknowledge the \fundingAgency{Czech Science Foundation} grant No. \fundingNumber{GX21-06825X}, and D.L. the internal grant of \fundingAgency{Silesian University} \fundingNumber{SGS/27/2022}. The authors were supported by the \fundingAgency{ESF} projects No. \fundingNumber{CZ.02.2.69/0.0/0.0/18-054/0014696}. A.Y. acknowledges the institutional support from \fundingAgency{Astronomical Institute of the Czech Academy of Sciences} with \fundingNumber{RVO:6798581}. This research was supported in part by the Polish \fundingAgency{NCN} grant No. \fundingNumber{2019/33/B/ST9/01564}. The computations in this work were supported by the \fundingAgency{PLGrid Infrastructure} through which access to the Prometheus supercomputer, located at ACK Cyfronet AGH in Krak\'{o}w, was provided.

\bibliography{puffy.bib}

\end{document}